\documentclass[sn-mathphys,Numbered]{sn-jnl}

\usepackage{graphicx}%
\usepackage{multirow}%
\usepackage{amsmath,amssymb,amsfonts}%
\usepackage{amsthm}%
\usepackage{mathrsfs}%
\usepackage[title]{appendix}%
\usepackage{xcolor}%
\usepackage{textcomp}%
\usepackage{manyfoot}%
\usepackage{booktabs}%
\usepackage{listings}%
\usepackage{lmodern}
\usepackage[british]{babel}

\newcommand{\myfigref}[1]{Fig.~\ref{#1}}
\newcommand{\myeqref}[1]{Eq.~\ref{#1}}
\newcommand{\mytabref}[1]{Tab.~\ref{#1}}
\newcommand{\mysecref}[1]{Sec.~\ref{#1}}
\newcommand{\gr}{$\gamma$-ray }
\newcommand{\grn}{$\gamma$-ray}
\newcommand{\grs}{$\gamma$-rays }

\begin{document}

\title{A computer-vision aided Compton-imaging system for radioactive waste characterization and decommissioning of nuclear power plants}

\author*[1]{\fnm{V{\'i}ctor} \sur{Babiano-Su{\'a}rez}}\email{vbabiano@ific.uv.es}
\author[2]{\fnm{Javier} \sur{Balibrea-Correa}}\email{javier.balibrea@ific.uv.es}
\author[2]{\fnm{Ion} \sur{Ladarescu}}\email{Ion.Ladarescu@ific.uv.es}
\author[2]{\fnm{Jorge} \sur{Lerendegui-Marco}}\email{Jorge.Lerendegui@ific.uv.es}
\author[3]{\fnm{Jos{\'e} Luis} \sur{Legan{\'e}s-Nieto}}\email{JLEN@enresa.es}
\author[2]{\fnm{C{\'e}sar} \sur{Domingo-Pardo}}\email{Cesar.Domingo@ific.uv.es}

\affil[1]{\orgname{Universitat de València}, \orgaddress{\street{Blasco Ibañez}}, \city{València}, \postcode{46010}, \state{Comunitat Valenciana}, \country{Spain}}

\affil[2]{\orgdiv{Instituto de F{\'\i}sica Corpuscular}, \orgname{CSIC-University of Valencia}, \orgaddress{\street{Catedr{\'a}tico Jos{\'e Beltr{\'a}n}}, \city{Paterna}, \postcode{46980}, \state{Comunitat Valenciana}, \country{Spain}}}

\affil[3]{\orgname{ENRESA}, \orgaddress{\street{Emilio Vargas}, \city{Madrid}, \postcode{28043}, \state{Madrid}, \country{Spain}}}

\abstract{Nuclear energy production is inherently tied to the management and disposal of radioactive waste. Enhancing classification and monitoring tools is therefore crucial, with significant socioeconomic implications. This paper reports on the applicability and performance of a  high-efficiency, cost-effective and portable Compton camera for detecting and visualizing low- and medium-level radioactive waste from the decommissioning and regular operation of nuclear power plants. The results demonstrate the good performance of Compton imaging for this type of application, both in terms of image resolution and reduced measuring time. A technical readiness level of TRL7 has been thus achieved with this system prototype, as demonstrated with dedicated field measurements carried out at the radioactive-waste disposal plant of El Cabril (Spain) utilizing a pluarility of radioactive-waste drums from decomissioned nuclear power plants. The performance of the system has been enhanced by means of computer-vision techniques in combination with advanced Compton-image reconstruction algorithms based on Maximum-Likelihood Expectation Maximization. Finally, we also show the feasibility of 3D tomographic reconstruction from a series of relatively short measurements around the objects of interest. The potential of this imaging system to enhance nuclear waste management makes it a promising innovation for the nuclear industry.}


\maketitle

\section{Introduction}\label{sec:intro}
The growth of electricity production based primarily on uranium fission has been steady since the 1950s, with approximately 440 nuclear power reactors currently operating in 32 countries, boasting a combined capacity of about 390 GWe~\cite{WNA2022}. The main advantage of using this technology for electricity production is the drastic reduction of harmful emissions~\cite{IPCC2014,IAEA2020}. In contrast, the generation of electricity from a typical 1000 MW nuclear power station produces around three cubic meters of vitrified high-level waste per year~\cite{WNAweb}. 

The production and subsequent disposal of radioactive waste is the main drawback of nuclear-energy-based electricity production. Consequently, some countries, such as Spain, Switzerland or Germany, have opted to phase out nuclear power. This leads to the decommissioning of the existing nuclear plants and the management of all associated nuclear residues and activated structural materials. Both regular operation and decommissioning of nuclear plants constitute the primary sources of radioactive waste that must be classified and safely disposed.

The classification of radioactive material mainly depends on its activity concentration\footnote{Except for the very-short-lived waste (VSLW), which can generate high-level activity but does not require special disposal facilities, since it will become regular waste within a short time period of a few years.}. From highest to lowest activity, the classification would be: high-, intermediate-, low- and very-low-level waste (HLW, ILW, LLW and VLLW). The first two categories primarily consist of spent fuel. On the contrary, LLW and VLLW, such as protective shoe covers and clothing, wiping rags, mops, filters, reactor water treatment waste, equipment and tools, etc, mainly originate from reactor operations and typically exhibit radioactivity just above background levels found in nature~\cite{IEA2009}. 

The disposal of radioactive waste varies according to the classification criteria exposed above. Therefore, VLLW and LLW are commonly disposed of in near-surface facilities until they decay sufficiently to be considered common waste, while ILW and HLW must be disposed long time in underground facilities specially designed for this purpose. In the case of HLW, special treatment is necessary due to the significant amount of heat produced by the radioactive decay process, in comparison to ILW. 

Given the cost, duration and limited space for radioactive waste disposal, effective management is essential to minimize the amount of material classified, which necessitates dedicated disposal centers as explained earlier. This management process involves not only measuring the activity but also assessing the homogeneity and distribution of contaminants throughout the materials. To address these challenges, research groups worldwide are actively developing and applying new detection devices and techniques~\cite{ Caballero2018,Lyoussi2000, Tortajada2023,Mahon2019,Perot2018,Haudebourg2015}. However, one of the limitations of most detection systems is the rather low detection efficiency and the relatively long characterization times. An alternative to overcome such limiations is to use electronically collimated (Compton) cameras, which can provide a large field of view (FOV), detection sensitivity and portability~\cite{Wahl15,Parajuli2022, Yao2022, Sato2020}.

In this work, we present the results of several field measurements for radioactive waste characterization using a RGB-camera coupled to a high-efficiency $\gamma$-ray Compton imager, thereby enabling hybrid \gr and visible-image reconstruction. 

The hardware elements of the detection system itself are detailed in \mysecref{sec:setup}. The \gr imager provides a graphical representation of the spatial distribution of the radioactive waste or source by means of the Compton imaging technique, which is explained later in~\mysecref{sec:technique}.  Field-measurements were carried out at the disposal plant of El Cabril (ENRESA), thereby utilizing the methodology reported in~\mysecref{sec:experiment}. \mysecref{subsec:singlebarrel} describes the baseline scenario for demonstrating the technique, consisting of the measurement and analysis of one single barrel, which we also use to introduce the computer-vision technique implemented in this work (\mysecref{subsubsec:cv}). After validating the full image-reconstruction technique we apply it to a more complex scenario consisting of an array or ensemble of drums with different activity levels and configurations. Finally,~\mysecref{sec:conclusions} provides a summary of the main conclusions and the future perspective.

\section{Hybrid imaging system and ancillary equipment}\label{sec:setup}
\begin{figure}[ht]%
	\centering
	\includegraphics[width=0.42\textwidth]{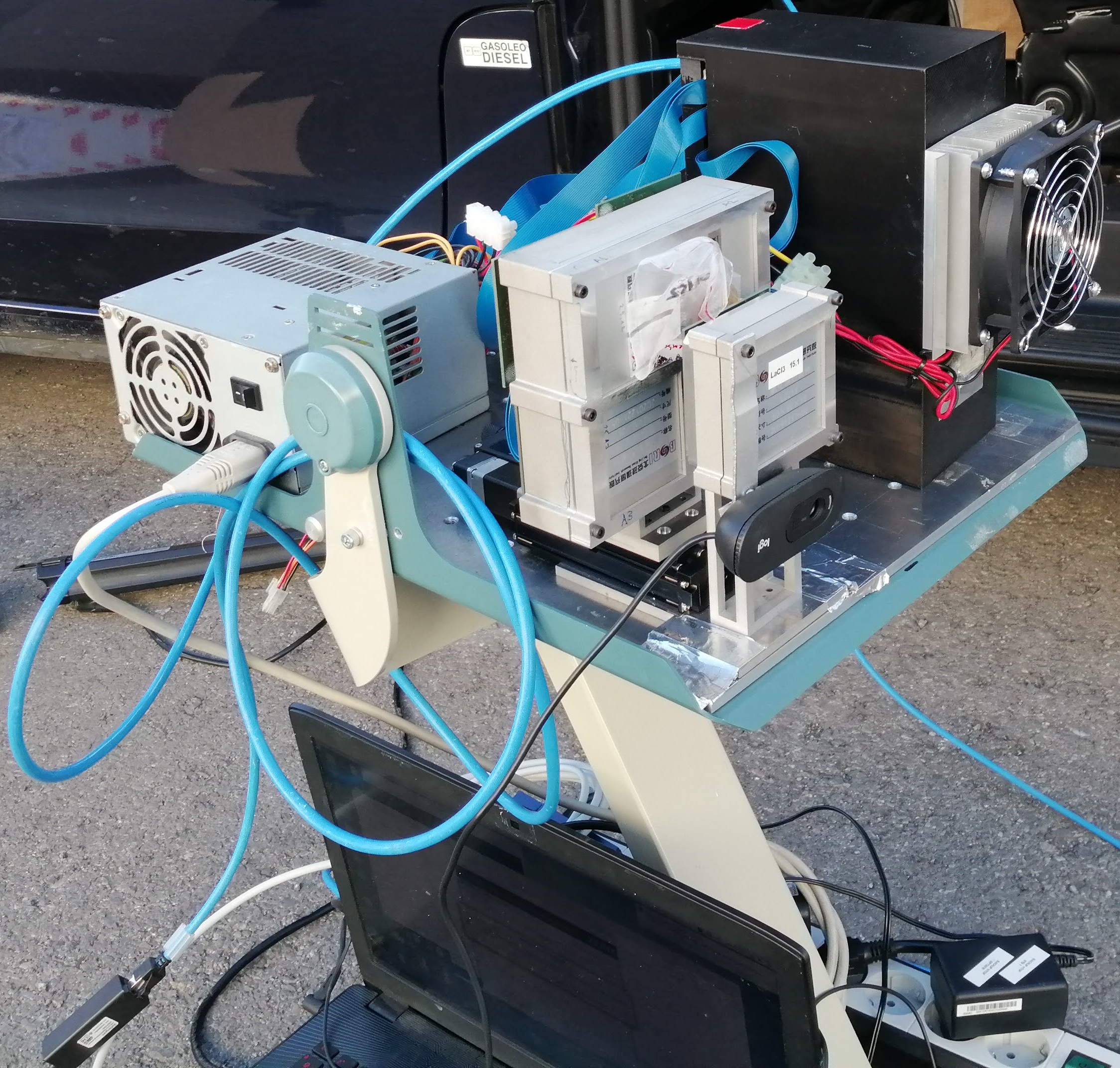}
	\caption{Photograph of the detection system mounted on a portable trolley. RGB-camera is attached to the front detector and directly connected to the laptop underneath.}
	\label{fig:ited}
\end{figure}

\myfigref{fig:ited} shows a picture of the assembled imaging prototype. The \gr imager itself was originally developed for nuclear astrophysics research~\cite{HYMNS}. One of its most prominent aspects is its large efficiency, which is related to its design~\cite{Babiano2020dec} that comprises an absorber (rear) detector with four times the area of the (front) scatter detector. Thus, the four-fold absorber acts as a large umbrella registering a large portion of the $\gamma$-rays scattered in the front detector. As reported in~\cite{Babiano2019}, this feature allows one to increase the number of scatter-absorber coincidence events for Compton imaging, thus remarkably enhancing detection efficiency with respect to conventional Compton-camera designs. This enhanced sensitivity becomes of particular relevance for applications requiring short measuring times or online monitoring. In addition, a positioning drive (Zaber Technologies Inc. LRT1500AL) allows one to remotely control the distance between absorber and scatter planes, thus enabling a trade-off between efficiency and image resolution~\cite{Domingo2015, Babiano2020dec}.

Computer vision was achieved by means of an inexpensive RGB camera (Logitech HD C270)  attached to the front part of the Compton imager in a fixed position and covering the front field of view of the latter. The RGB camera has a diagonal field of view (dFOV) of 55$^{\circ}$ and a resolution of 0.9 Mpixels. The camera was connected via USB-A to a portable computer, which was also used for data acquisition and slow control operations. The laptop (ASUS GL552V) was based on an Intel Core i7-6700HQ at 2.6 GHz and the Graphics Processing Unit (GPU) NVIDIA GeForce GTX 950M. The full system was mounted on a small trolley, which allowed the operator to position it around the object of interest.

As described in~\mysecref{sec:technique}, Compton imaging requires of the simultaneous measurement of the 3D-coordinates of the \gr hit position and its deposited energy in both detection layers. To this aim Position Sensitive Detectors (PSDs) are employed. One PSD is used as scatter detector, while an array of four PSD units is utilized as absorber. Each PSD comprises a LaCl$_3$(Ce) monolithic crystal, with a base size of 50$\times$50 mm$^2$ and a thickness of 10 mm (25 mm) for the scatterer (absorber) detector. Scintillation photons generated in each crystal are readout by means of a Silicon Photomultiplier (SiPM) SensL (ArrayJ-60035-65P-PCB), which covers a surface of 50$\times$50 mm$^2$ with a segmentation of 8$\times$8 pixels. The 3D-coordinates of the \gr hit in each crystal volume are obtained by means of advanced positioning algorithms specifically developed for these PSDs~\cite{Babiano2019, Balibrea2021}. Analogue signals from each SiPM are digitized by means of Application-Specific Integrated Circuit (ASIC)-based (TOFPET2) readout electronics developed by PETsys~\cite{Rolo2013, Francesco2016}. The performance of the ASICs is significantly affected by variations of the temperature, which were minimized during the measurements by means of 20$\times$20 mm$^2$ Peltier cells (FPH1-7106NC) thermally coupled to each ASIC, as described in Ref.\cite{Babiano2020dec}. Still, gain variations in this field measurements were significantly larger than in laboratory conditions~\cite{Olleros2018, Babiano2019, Babiano2021}. In order to mitigate this effect the value of the temperature was registered in the neighbourhood of each ASIC in small time intervals as a function of the measuring time and a numerical correction for the gain was applied in the offline analysis of the data. On average, an energy resolution of 7\% \textsc{fwhm} at 662 keV was obtained, while spatial resolutions were of $\sim$2 mm \textsc{fwhm} in the transversal crystal (XY plane) and $\sim$5~mm \textsc{fwhm} in the orthogonal Z-coordinate or depth of interaction\cite{Balibrea2021}. After several tests for this particular application, the distance between scatter- and absorber-modules was set to 35 mm, which provided a convenient balance between resolution and efficiency for the present measurements~\cite{Domingo2015, Babiano2020dec}. 

\section{Field measurements at the disposal plant of El Cabril}\label{sec:experiment}
Field measurements were conducted at the disposal facility of El Cabril, which is located in Córdoba, Spain~\cite{CabrilWeb}. This facility is governed by the national Waste Management Organization (WMO) for the management of radioactive waste, ENRESA~\cite{EnresaWeb}. The disposal plant includes disposal cells specifically designed to safely isolate and contain VLLW, LLW and ILW generated at different national nuclear-power plants. In thi study, several drums filled with radioactive material from the decommissioning and regular operation of the power plant of Santa María de Garoña (GR), a BWR with 460 MW, and Ascó (AS), a pressurized-water reactor (PWR) of 1032 MW, were measured. The waste was contained in five different drums made of steel, each with dimensions of 28 cm radius and 85 cm height and a maximum capacity of 700 kg. Barrels labeled with AS are filled with a light compound (cellulose). Their content has been pressed and compacted and it is expected to be heterogeneous. On the other hand, drums marked with GR contain chemicals and other liquids that have been mixed with cement and solidified, they are expected to be less heterogeneous than AS barrels.

The drums measured in this work were previously characterized by the producer and subsequently validated by ENRESA following the standard procedure based on dose-rate measurements on contact surface. To this aim several measurements around the drum are carried out. For the final quoted contact dose rate (CDR), the average of several measurements performed at half height is usually taken. This methodology takes into account that different CDRs can be measured for the same activity depending on the measurement point. Obviously, this procedure only provides accurate results for rather homogeneous activity distributions in the barrel. Additionally, with the conventional approach an assessment of the activity content is made, thereby considering attenuation factors which depend upon the nature of the residues, the inner shielding material (cellulose or concrete), and density (measured weight and volume). In general, quoted activities normally correspond to over-estimated or conservative quantities.

Most of the activity present in the drums arises from the natural decay of $^{60}$Co (t$_{1/2} = 5.27$ y). The latter is artificially produced after neutron activation of steel structures in the reactor. $^{60}$Co decays to the stable isotope $^{60}$Ni, thereby emitting a $\beta$-particle, an antineutrino and two \grs from excited states in $^{60}$Ni with characteristic energies of 1.173 MeV and 1.332 MeV. These two \gr lines showed up prominently in the measured spectra and thus they were used in this work for estimating the activity of each drum and for visualizing their spatial distributions, as it will be discussed later.

\mytabref{tab:activityBarrels} shows the activity of $^{60}$Co estimated with the conventional methodology for five different drums utilized in this work. In the case of AS barrels, the activity shown in this table was obtained from the average value of the CDR quantities measured at half-height, as explained above. In addition, the value quoted for the AS28298 drum was evaluated including an additional independent measurement carried out with a high-efficiency and high-resolution LaBr$_3$ detector. On the contrary, small but representative samples of the initially liquid waste that forms GR barrels are taken before conditioned with cement for gamma spectrometry. The latter result is then extrapolated taking into account the mass of waste contained in the drum and assuming an homogeneous distribution of the activity across the barrel. In both cases conversion factors are applied to calculate the final activity depending on the isotopic content and barrel composition. For the sake of clarity, the listed quantities have been recalculated for the date of the reported measurements (February 2nd, 2022) utilizing the values obtained by the time of the characterization and the accurately known value for the half-live of $^{60}$Co. 

\begin{table}[h!]
    \centering
    \begin{tabular}{c c c}
        \hline
        Barrel ID & Total Estimated Activity Content & Effective (External) Activity\\
        \hline
        AS27954 & 284  & 284 \\
        AS28298 & 386  & 386 \\ 
        AS29448 & 38  & 38 \\
        GR15748 & 329  & 109 \\
        GR36256 & 341  & 112 \\
        \hline
    \end{tabular}
    \caption{Activity (in MBq) due to the decay of $^{60}$Co in the five drums used in this work. The second column indicates the overall $^{60}$Co activity estimated for each drum. The last column shows the external $^{60}$Co activity, estimated from the CDR measurements.}
    \label{tab:activityBarrels} 
\end{table}

One of the main uncertainties in the values quoted in~\mytabref{tab:activityBarrels} with the conventional methodology arises from the rather unknown internal distribution of mass and activity inside the drums. Additionally, for both types of barrels, in order to calculate an average activity value a strong assumption is made, where a full homogeneity of the activity distribution inside the barrel is assumed. Such hypothesis is well suited for GR barrels (because of the homogenized content). For AS barrels, which contain compacted solid heterogenous material, the full-homogeneity approximation is more error prone~\cite{Leganes2019}. 
Additionally, in the case of GR barrels, both the heterogeneous content and the shielding effect caused by their massive concrete filling introduce further uncertainties in the conventional methodology. Approximately only one third of the total activity assigned to these barrels can be effectively measured on the surface, as it is shown in the third column of~\mytabref{tab:activityBarrels}.

In this work several measurements were carried out with the hybrid imaging system described in detail in~\mysecref{sec:setup} utilizing five available drums of AS and GR origin. In a first run, a single unit (AS29448) was measured with the hybrid imaging system from different poses. This first measurement was intended to validate the methodology and characterization elements. The high \gr detection efficiency of the imager enabled sufficient counting statistics for measurements of only two minutes (120 s) in each pose, thus requiring in total less than 15 minutes for the full characterization of one barrel. The single-barrel analysis and characterization methodology is presented below in~\mysecref{subsec:singlebarrel}. After validating the methodology for one single drum a series of measurements were carried out with two different configurations of the five available units. The analysis methodology and results are reported in \mysecref{subsec:barrelarrangement}. 

\subsection{Characterization of one single barrel}\label{subsec:singlebarrel}
The system presented in~\mysecref{sec:setup} is intended to generate a 2D image or heat map depicting the spatial distribution of the $\gamma$-radiation overlaid onto a picture captured by the RGB camera. The heat map is provided in a color scale calibrated in activity units (Bq). To this aim the different techniques described below are employed. First, a Maximum-Likelihood-based algorithm is used to reconstruct the heat map as depicted in~\mysecref{sec:technique}. Second, computer vision techniques (\mysecref{subsubsec:cv}) allow one to consistently combine the radiation heat map with the RGB picture. Third, as detailed in \mysecref{subsubsec:activity}, Monte Carlo (MC) simulations are conducted in order to calibrate the response of the Compton camera to the detected radiation, thereby enabling an estimation of the measured activity. Finally, measuring from different perspectives around the object(s) of interest enables one to visualize the 3D distribution of the \gr radiation, as it will be explained in \mysecref{subsubsec:3D}.

\subsubsection{Maximum-Likelihood based image reconstruction algorithm}\label{sec:technique}
The Compton technique normally makes use of two detection planes that operate in time coincidence. Therefore, \grs can be scattered in the first plane, called scatterer, for then being fully absorbed in the second plane, called absorber. Information on the energy, position, and time of \gr interactions in the two detection planes is required. The interaction positions of the same \gr in both detection planes draw the axis of a cone whose aperture angle $\theta$ is given by the Compton formula expressed in~\myeqref{eq:compton}. There, $m_e$ is the mass of the electron in rest, $c$ the speed of light, $E_2$ the energy deposited in the second interaction, and $E_\gamma$ the energy of the \grn, which is assumed to be equal to the sum of energies deposited in both detection planes $E_\gamma = E_1 + E_2$. 

\begin{equation} \label{eq:compton}
    \theta = arccos \left [ 1 - m_e c^2 \left ( \frac{1}{E_2} - \frac{1}{E_\gamma} \right ) \right ]
\end{equation}

In a first approach, the image of a \gr source can be obtained from the superposition of the Compton cones at a plane placed at the source distance, oriented in parallel to the detection planes~\cite{Wilderman1998}. The resolution and the signal-to-background ratio of the back-projected image are rather limited. The performance can be further improved by applying an iterative method based on Maximum Likelihood Expectation Maximization~\cite{Richardson72,Lucy74}, and in particular for List Mode Likelihood Expectation Maximization (LMLEM)~\cite{Barrett:97} for Compton Cameras~\cite{Wilderman:98-1,Wilderman:98-2}. Assuming a pixelated object described by the vector $\vec{\lambda}$, the LMLEM algorithm takes a compact form~\cite{Barrett:97,Wilderman:98-1,Wilderman:98-2}. During each iteration $l$, the pixel $j$ intensity value ($\lambda_{j}$) is updated as follows

\begin{equation}
\lambda^{l+1}_{j}=\frac{\lambda^{l}_{j}}{s_{j}}\sum_{i}{\frac{t_{ij}}{\sum_{k}t_{ik}\lambda^{l}_{k}}},
\end{equation}

\noindent where $s_{j}$ represents the system sensitivity for the pixel $j$ and $t_{ij}$ is the transition matrix for one individual event $i$ at the pixel $j$. The latter can be calculated utilizing the back-projection algorithm~\cite{Wilderman1998} using a spread function to account for the detector angular resolution~\cite{Wilderman:98-1}. In this work, a spread function with a constant parameter $\sigma$~\cite{Wilderman:98-2} was used under the approximation of mono-energetic $\gamma$-rays, regardless of the particular \gr energy. Additionally, we have adopted the far FOV approximation, where the distance between the detector and the image plane is large and $s_{j}$ can be considered constant. As discussed in~\cite{Wilderman:98-2}, a more sophisticated image reconstruction based on an accurate modelling of both, $s_{j}$ and $t_{ij}$, can be developed for more complex scenarios. 

\subsubsection{Computer vision}\label{subsubsec:cv}
The RGB camera embedded in the detection system serves the purpose of capturing the scene being measured by the Compton camera. For this goal, the camera was set-up with its sensor plane parallel to the detection planes of the Compton device and thus the visual field of view was oriented in the same direction as the \gr imager. The resulting picture is analyzed using the computer-vision tools described below, which allow one to infer the location and distance of the barrel ensemble, as well as the precise location of every drum within the image with respect to the imaging system. These tasks are accomplished using two different techniques. 

The first technique implemented was based on the use of fiducial markers, in a similar fashion as described in previous works \cite{Caballero2018,Tortajada2023}. To this aim, fiducial markers were attached to the radioactive barrels in order to obtain the 3D coordinates of the detection system relative to them. Detecting these fiducial markers within the RGB image provides information on the distance (Z coordinate), which becomes crucial for determining the correct image plane where the Compton image is reconstructed. Furthermore, the size in the transversal plane (XY), in pixels, is utilized to scale the Compton and RGB images properly through a transformation factor that relates both units, meters and pixels. For this purpose, we developed an algorithm based on the Aruco C++ library~\cite{Aruco2015,Aruco2018}, which was chosen due to its robustness and straight forward implementation in our C++ based software. The geometric calibration of the RGB- and \gr cameras was accomplished by means of dedicated laboratory measurements using a $^{22}$Na source in combination with fiducial markers, and following the common procedure described in previous works \cite{Caballero2018, Tortajada2023}. 
\begin{figure}[htbp!]%
    \centering
    \includegraphics[width=\textwidth]{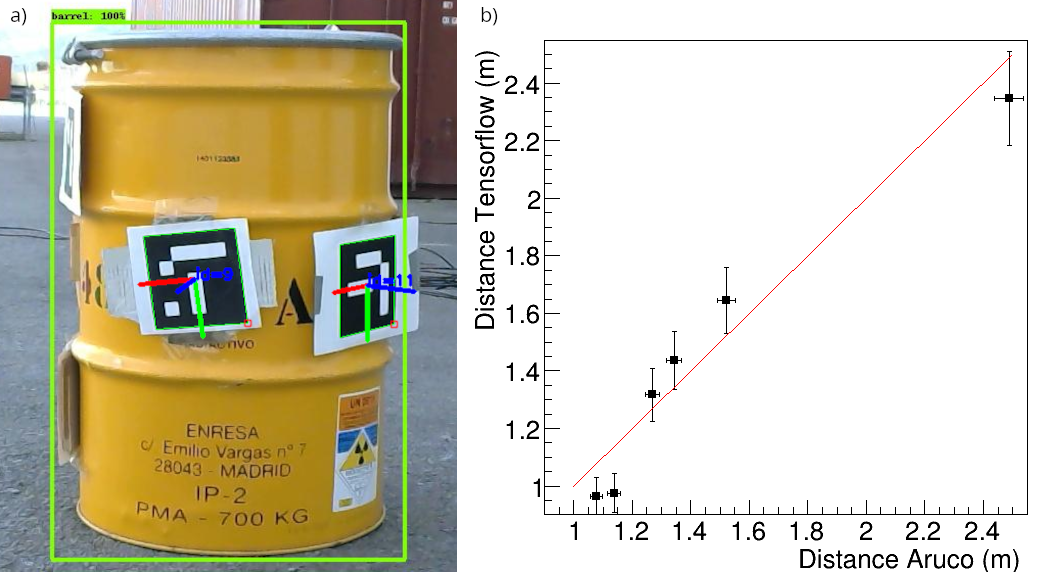}
    \caption{a) Photograph of the isolated drum from position 4 including the detection of Arucos and barrels. b) Black points compare distances measured using Tensorflow and fiducial markers (Aruco) for the six posses. Red line shows the ideal situation.}\label{fig:aruco}
\end{figure}
An alternative methodology based on the use of Artificial Intelligence (AI) was also implemented. The use of differential programming techniques enables the individual identification of each barrel, yielding their X and Y coordinates in the image. This spatial information not only aids in resolving the Compton image within the region occupied by the drums but also offers an alternative method for getting the Z distance without need of fiducial markers. In practical terms, avoiding the use of fiducial markers has a twofold added value because it allows reducing operator radiation dose and it also facilitates the full characterization procedure. Since the initial X and Y dimensions of the drums are known, determining their size in the image allows one to determine their distance relative to the detection system. In order to accomplish this task, we employ a custom-trained model based on Tensorflow~\cite{Tensorflow2015}, which is an open-source library developed by Google for machine learning (ML) tasks, such as object detection. The performance of this methodology is demonstrated and validated in \myfigref{fig:aruco}-a, which shows a picture taken with the RGB camera after being simultaneously analyzed both with the conventional methodology based on fiducial markes and with the aforementioned computer-vision code. Two fiducial markers in the picture are detected, enclosed within green squares, and a 3D reference system is assigned to each of them. Additionally, the drum is also identified by the AI model, as shown in the figure by the green rectangle enclosing it. A score value indicates the confidence level of the detection (100\%). Both fiducial marker and barrel identification occurs almost instantaneously (at a ms scale), thus enabling the potential implementation of these solutions for real-time applications in the future.

\myfigref{fig:aruco}-b shows the correlation between the Z distances for the six measured perspectives obtained using Tensorflow and the fiducial markers. From this comparison a relative uncertainty of around 7\% can be assigned to the values obtained with the new technique. These discrepancies are mainly due to the precision of the latter in delineating the space occupied by the drum. One could improve this accuracy further by means of a more refined segmentation model that allows one to infer the exact boundaries of the detected barrels. However, for the purpose of the present work we regard the obtained result as satisfactory and further enhancements, at this stage, have not been undertaken.

\subsubsection{Real-time determination of the apparent surface activity}\label{subsubsec:activity}
In order to improve the regular procedure for classification and management of radioactive waste a rapid assessment of the activity level is crucial. Because a Compton imaging system has no mechanical collimation, such information  can be obtained with a high efficiency (or short measuring time) simply by using the single-energy spectra of each PSD embedded in the Compton camera. The precise value for the activity can be derived during the inspection from measured or characterized quantities, such as the measurement duration, environmental \gr background, and detection system efficiency, in addition to the nature (homogeneity and density) of the radioactive material being analyzed. In this work we have implemented the method outlined in~\cite{Ridikas2005}, which comprehensively considers all these variables through a combination of spectrometric data and Monte Carlo (MC) simulations. With the present setup, the simultaneous five-fold independent measurement of the activity (by the scatter and four absorber PSDs), provides a rather reliable assessment from the systematics and statistical accuracy point of view.
\begin{figure}[ht]
    \centering
    \includegraphics[width=0.95\textwidth]{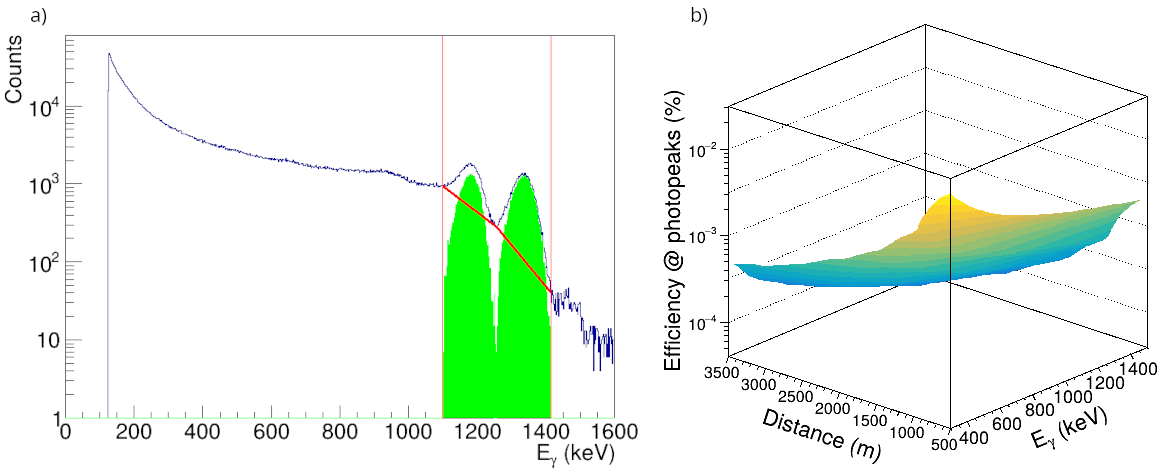}
    \caption{a) Spectrum of deposited energy measured with one PSD of the Compton camera (blue line). Red dashed lines shows the analysis selection for full-energy events while the red solid line corresponds to the estimated background.   b) Efficiency of the i-TED Scatterer as a function of the distance and the energy of the photopeak.}\label{fig:activity}
\end{figure}

\myfigref{fig:activity}-a shows an example for a \gr spectrum measured with one of the PSDs in the camera. The energy resolution of the LaCl$_3$(Ce) monolithic crystals allows one to distinguish the two photopeaks originating from the $^{60}$Co decay. A dedicated analysis software permits to identify these two \gr transitions (enclosed by two red dashed lines in the figure), while the natural background (red solid line) is estimated by means of the broken-line method~\cite{Celik2022} and polynomial forms of degree 1. Subtraction of this background yields the green-filled histogram, which represents the contribution of the $^{60}$Co decay. Since the measurement time is known and the approximate distance to the source is estimated as described in the previous section, the activity of the source can be determined from a previously determined efficiency for the detection system~\cite{Ridikas2005}.

The detection efficiency for each PSD was determined by means of MC simulations using the C++ based \textsc{Geant4} toolkit~\cite{Allison2016}, in a similar way as we did in a previous work~\cite{Caballero2018}. A realistic implementation of the Compton camera was included in the geometry package. In each simulation, 10$^8$ mono-energetic \grs with energies ranging from 300 keV to 1.5 MeV were emitted from an isotropic extended source positioned at distances spanning from 0.5 m to 3.5 m. A total of 140 simulations were conducted, covering 20 different gamma ray energies and 7 distances. These simulated data enable the derivation of a 2D surface (distance - energy) representing the efficiency at the photopeak for each PSD. \myfigref{fig:activity}-b displays this surface for the PSD that serves as scatter detector in the \gr camera. Considering the size of the drums under inspection, the spatial distribution of the \gr source was found to have a minor impact in the estimated efficiencies for distances larger than one meter. Even so, a radiative surface of approximately the size of the drum projection was simulated as a \gr source.

\begin{figure}[ht]%
    \centering
    \includegraphics[width=\textwidth]{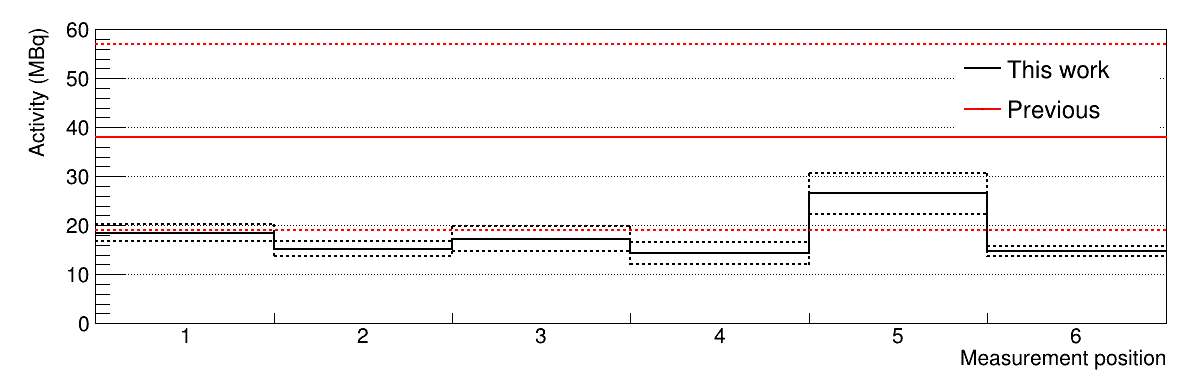}
    \caption{Activity of the individual barrel measured in this work from the 6 different positions and compared to the previous measurement provided by ENRESA.}\label{fig:activityOne}
\end{figure}

Finally, the apparent surface activity can be derived from each PSD, or an average value of them, by taking into account the net number of counts in the photopeaks and the efficiency obtained from the MC simulation at specific distances and energy values of the source (\myfigref{fig:activity}-b). The result obtained for the single drum study (AS29448) is shown in~\myfigref{fig:activityOne} for the six measured poses together with their corresponding root mean-square (\textsc{rms}) deviation. The latter reflects mainly systematic uncertainties arising from distance estimates and approximations in detector response or efficiency. As it can be observed, most of the measured poses yield consistent values for the estimated activity, ranging between 14.3 MBq and 18.5 MBq, with a \textsc{rms} deviation of $\pm$15\%. This result is in agreement with the previous value of 38(19) MBq obtained with contact dose-rate measurements (red line in~\myfigref{fig:activityOne}). One should also consider the accuracy of a factor of 2-3, which is expected for the implemented method, as reported in Ref.\cite{Ridikas2005}. 

It is particularly interesting to discuss the results obtained for pose 5, which essentially doubles the value obtained for the other poses. Indeed, for pose 5 an apparent surface activity of 27(4) MBq is obtained. This apparent discrepancy points to a significant inhomogeneity in the activity or in the density distribution within the analyzed barrel. Nevertheless, it is worth noting that the value provided by ENRESA for this drum (38 MBq in \mytabref{tab:activityBarrels}) corresponds to a rather conservative approach, estimated from a series of dose-rate surface measurements around the barrel. That result, which is about 40\% higher than the maximum activity obtained in this work (pose 5), still falls within the expected systematic uncertainty of a factor 2-3 for the technique implemented here~\cite{Ridikas2005}.

\subsubsection{Hybrid \texorpdfstring{\grn}{gamma-ray} and visible 2D image}\label{subsubsec:2Dimage}
The \gr heat map of the measured drum or container, obtained via the Compton imaging algoritm described in \mysecref{sec:technique} and calibrated in activity units (\mysecref{subsubsec:activity}), can be overlaid onto the visible image captured by the RGB camera aided with the Computer vision techniques described in \mysecref{subsubsec:cv}. This process delivers an hybrid image, which combines both spatial and radiation (activity) information. This procedure is detailed in the following.

The Compton image is generated by selecting the two peaks corresponding to full-energy deposition events from the two \gr transitions in the decay of $^{60}$Co (see~\myfigref{fig:activity}-a). At each coincidence event between both detection layers, the reconstructed 3D \gr hit positions and the deposited energies are used as input for the LMLEM algorithm (see \mysecref{sec:technique}).

\begin{figure}[!htbp]%
    \centering
    \includegraphics[width=\textwidth]{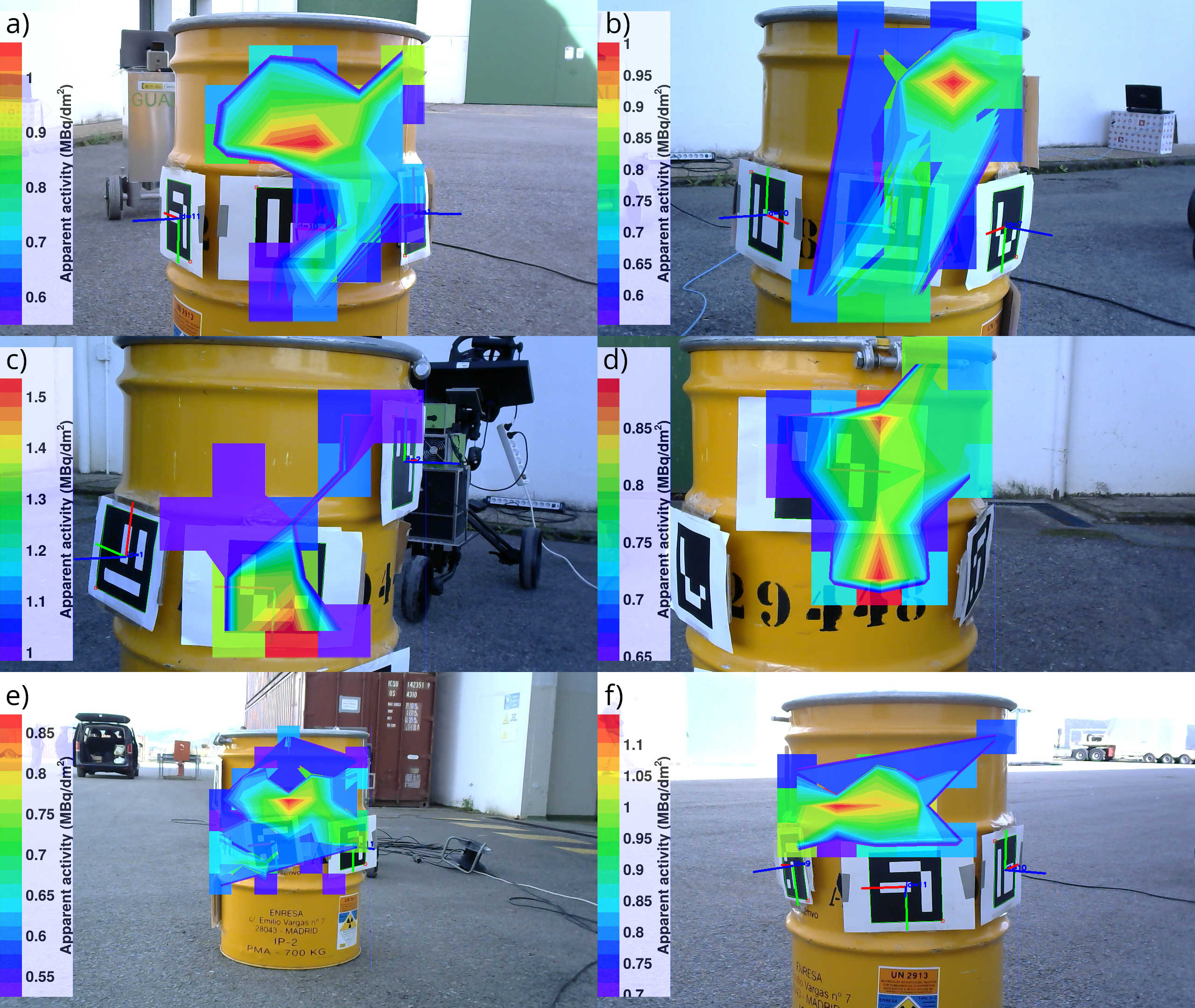}
    \caption{Hybrid \gr and visible 2D images for the six poses measured around the single drum.}\label{fig:barrel}
\end{figure}

The average activity determined as described in the previous section from the singles energy spectra of the individual PSDs is now assumed to be distributed over a flat surface following the Compton 2D image obtained with the LMLEM algorithm in that plane. This is a reasonable assumption, considering that all detected $^{60}$Co transitions occur in the front field-of-view of the $\gamma$-ray imager and the RGB-sensor, as it is the case by construction. In other circumstances, for example with the $\gamma$-ray source in the front and rear parts of the detector, a more sophisticated reconstruction algorithm would be required. The hybrid \gr and visual images obtained with this methodology are shown in \myfigref{fig:barrel} for the six poses around the drum. 

The resulting hybrid images illustrate that the activity is predominantly concentrated in the upper part of the drum, except for positions 3 and 4 (\myfigref{fig:barrel}-c and -d), where the maximum activity is shifted toward the central region of the barrel. The most active part of the bulk is situated in the top third fraction of the face captured in pose 5 (\myfigref{fig:barrel}-e). This observation is still visible for pose 6, where the main locus of activity is shifted towards the left.

\subsubsection{3D hybrid-image reconstruction}\label{subsubsec:3D}
Multiple perspectives enable one to obtain a 3D distribution of the \gr radiation using a tomographic approach. This method initiates by reconstructing 2D Compton images for several parallel image planes distributed along the Z direction. Reconstructed Compton cones for each event in each pose are then projected through the three-dimensional space discretized into voxels.

Combining the resulting histograms obtained from the six poses delivers a 3D back-projected image or heat map of the source activity distribution. The basic advantage of this simplified method is based on the fact that, despite of the large amount of CPU power required, the processing time is relatively short and, at least, smaller than the overall time required for the multiple pose acquisitions. 

In order to apply this 3D tomographic method it must be fulfilled for each pose that  
\begin{itemize}
    \item the detection system is pointing towards the drum,
 \item at least one fiducial marker is placed rather tangentially to the drum, and
    \item at least one fiducial marker from the following pose must be detected.
\end{itemize}
Under these assumptions one can compute the angle subtended between the poses, as well as the Z distances between the imaging system and the drum or the corresponding object of interest. It becomes then feasible to determine the position of the detection system in a global coordinate system shared across all the measured poses, which is necessary to combine all the measurements into a single 3D reconstructed image.

The geometrical transformations required to represent the $x' y' z'$ coordinates of the 3D histogram obtained at each position in the global coordinate system $x y z$ is given in~\myeqref{eq:3Dtransformation}, where $\theta$ represents the angle subtended between each posse and a reference one, and $r$ denotes the distance between the center of the drum and the detection system.
\begin{equation}\label{eq:3Dtransformation}
    \begin{aligned} 
        x &= (r-z')sin\theta + x'cos\theta \\
        y &= y' \\
        z &= x'sin\theta + (z'-r)cos\theta + r
    \end{aligned}    
\end{equation}

\begin{figure}[!htbp]%
    \centering
    \includegraphics[width=0.8\textwidth]{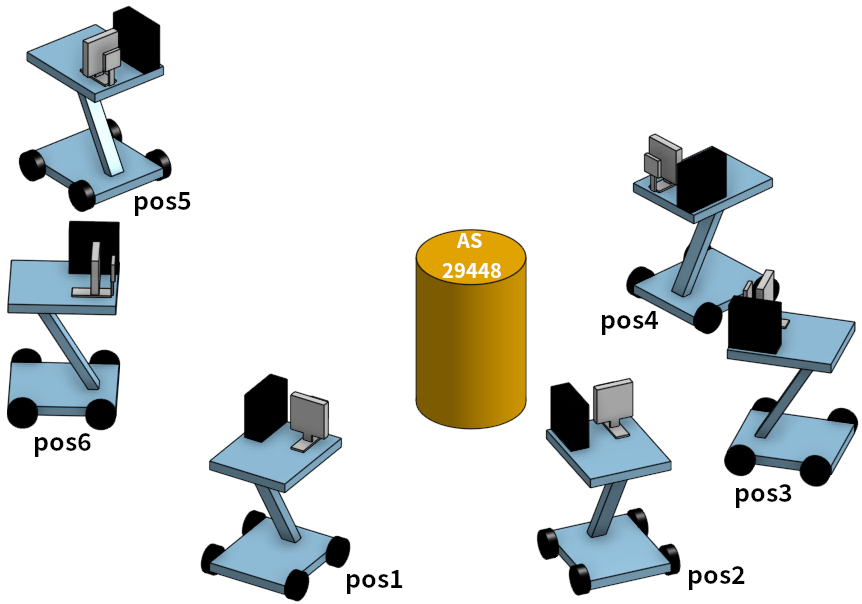}
    \caption{Schematic (CAD) view of the AS 29448 barrel and the measured poses at the corresponding distances and angles obtained via image-segmentation during the measurements.}\label{fig:cadBarrel}
\end{figure}

\begin{figure}[!htbp]%
    \centering
    \includegraphics[width=0.8\textwidth]{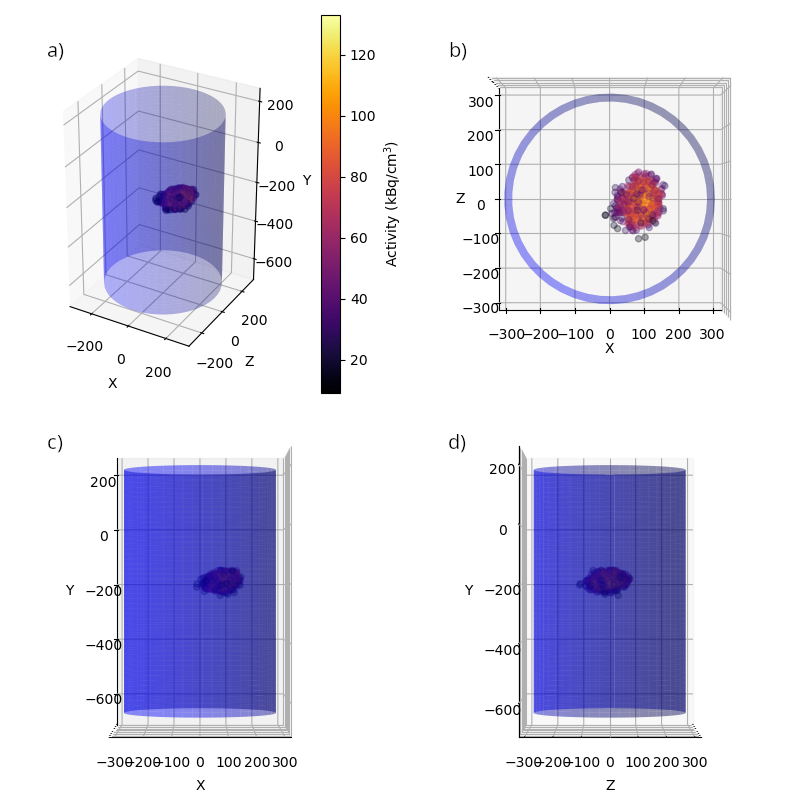}
    \caption{a) Cabinet view of the 3D distribution of \gr radiation of the simulated barrel. Top-, left and right-views are given in the following panels b), c) and d).}\label{fig:3DactivityMC}
\end{figure}

In order to validate this methodology on the basis of an accurately known and controlled scenario, a new set of MC simulations were carried out with \textsc{Geant4}. Both the drum with radioactive material and the detection system placed at six different measurement positions were realistically implemented, mirroring the setup of a virtual drum characterization. \myfigref{fig:cadBarrel} displays a CAD view of the drum and the measurement positions. Using this setup, $9.12 \times 10^{9}$ \grs with energies of 1173~keV and 1333~keV were simulated at each position, mimicking a 38 MBq $^{60}$Co source measured during 120 s. The modeled \gr source consisted of a radiative cube with a size of 10$\times$10$\times$10~cm$^3$ placed at the position (x=10,y=-20,z=0) cm inside the drum. \myfigref{fig:3DactivityMC} displays the 3D distribution of activity reconstructed within the 3D space, discretized using voxels of 1~cm$^3$ size. The 3D heatmap image shown in the latter figure is generated using a density scatter plot created with the Matplotlib library~\cite{Hunter2007} in Python. The color bar indicates activity units (kBq), distributed throughout the space, thereby taking into account the density of counts in the representation. Additionally, the geometry of the cylindrical drum is included in the figure to guide the eye in the interpretation of the result. The reconstructed image shows that the origin and extension of the \gr source are reasonably well determined. The location of the reconstructed \gr volume agrees within 5-10~cm for the $x$ and $z$ coordinates, while the $y$ coordinate remains unaltered. Several tests made with other reconstruction voxels showed that this accuracy can be mainly ascribed to the tomographic method and it is significantly influenced by the voxel size, with larger reconstruction voxels resulting in a degradation of this resolution but also in shorter processing times.

\begin{figure}[ht]%
    \centering
    \includegraphics[width=0.8\textwidth]{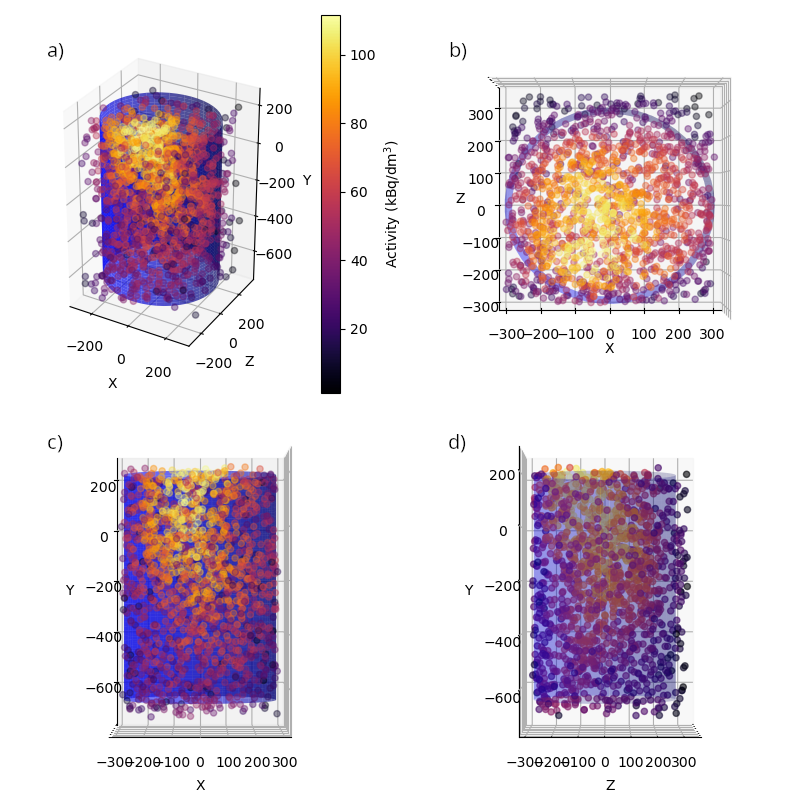}
    \caption{3D distribution of \gr radiation of the AS 29448 barrel.}\label{fig:3DactivityOne}
\end{figure}

After validating the methodology with simulated data the 3D reconstruction methodology was applied to the individual drum (AS29448) measured at El Cabril and used as example in the preceding sections. For this analysis 5 cm side voxels were chosen as a good tradeoff between image resolution and statistical accuracy. This voxel size is also in accordance with the position accuracy expected from the previous MC study. \myfigref{fig:3DactivityOne} displays the resulting 3D distribution of radioactivity, assuming also that the integral value of 27 MBq (see~\mysecref{subsubsec:activity}) was distributed along the drum volume, as discussed in~\mysecref{subsubsec:activity}. This result reveals that the activity is primarily concentrated within the negative $x$ and $z$ coordinates, predominantly around the top-half region of the barrel, which is consistent with the 2D images produced in~\mysecref{subsubsec:2Dimage}. The effective attenuation of the radiation across the barrel was included by means of MC simulations performed for both cellulose and concrete filling materials. The simulations yield effective attenuation coefficients of approximately 5 m$^{-1}$ and 10 m$^{-1}$, respectively, for the \gr energies utilized in the image reconstruction. In addition, a correction for the effect induced by the interference between the Compton arcs and the attenuation correction was also included in the analysis. Still, the results obtained for different attenuation coefficients do not affect significantly the reconstructed hybrid images, and in terms of activity determination they are within the uncertainties quoted for the method.

\subsection{Hybrid imaging of complex drum ensembles}\label{subsec:barrelarrangement}
\begin{figure}[ht]%
    \centering
    \includegraphics[width=0.55\textwidth]{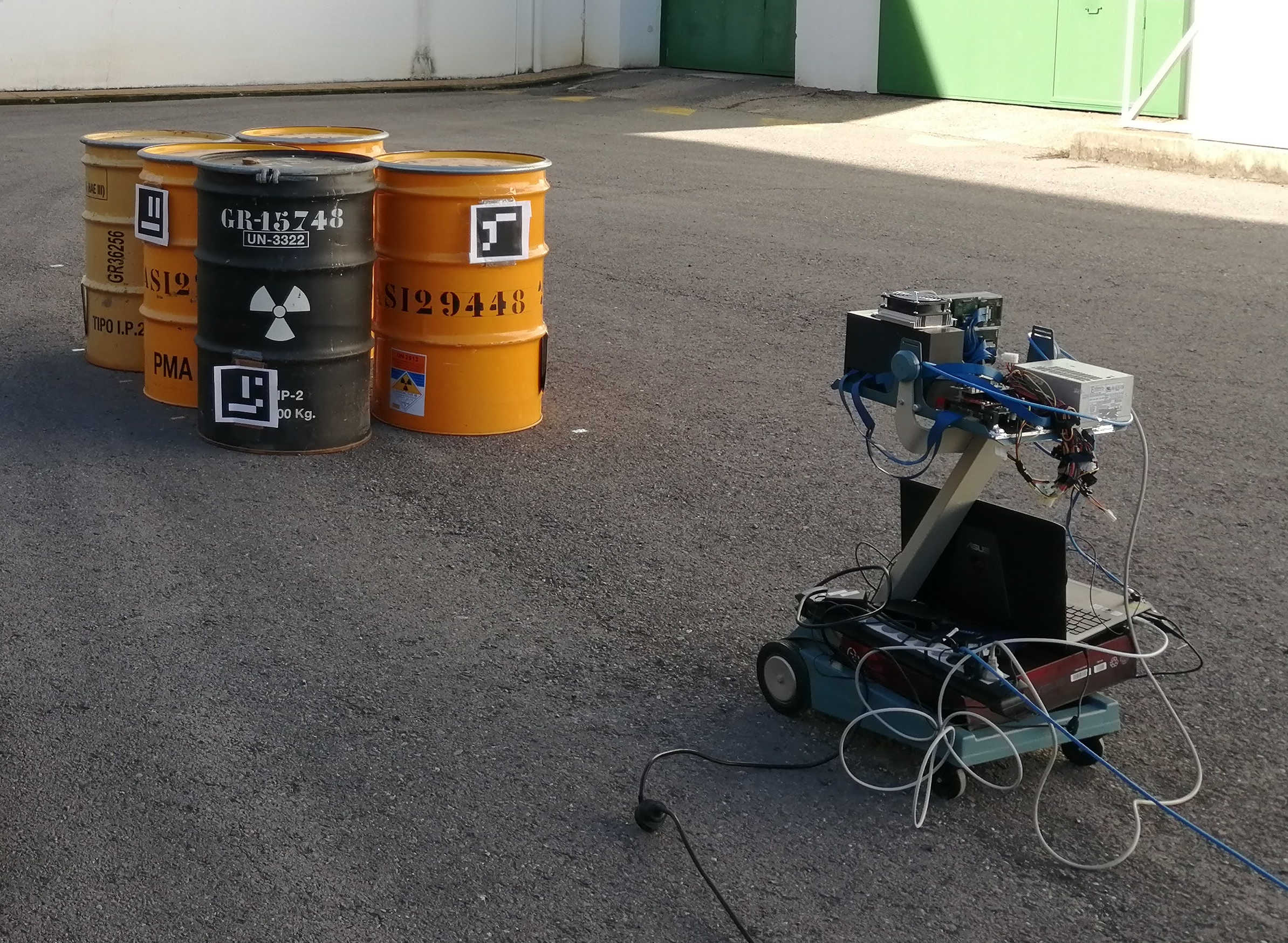}
    \caption{Photograph of the experimental setup used during the measurements at El Cabril.}\label{fig:setup}
\end{figure}

Empowered by the satisfactory results obtained for the image analysis of the individual barrel, the same methodology was extended to more complex scenarios, involving two different barrel arrangements, hereafter designated as configurations A and B. For each configuration, eight different perspectives or poses were measured, each of them lasting for 300 s. \myfigref{fig:setup} shows a photograph of the barrel arrangement A with the detection system positioned at one of the measured poses. Average distances between the drum ensemble and the imaging system were of about 200 cm. An accurate measurement of the distance by the operator was not required because a rather precise value was derived by exploiting the computer vision and image segmentation techniques, as described in~\mysecref{subsubsec:cv}.

\begin{figure}[ht]%
    \centering
    \includegraphics[width=\textwidth]{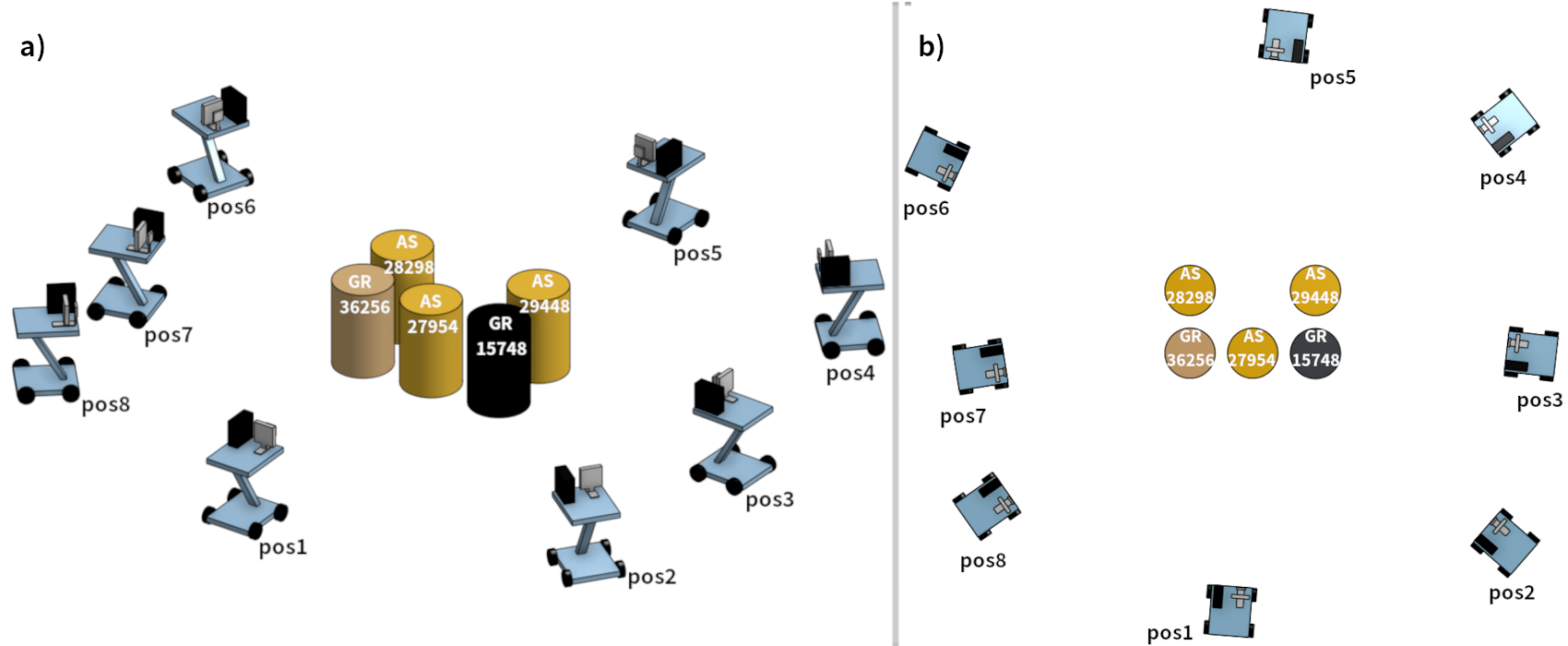}
    \caption{Schematic (CAD) views of the barrel configuration A together with the measured poses. a) Isometric- b) top-view.} 
    \label{fig:cadArrangementA}
\end{figure}

\begin{figure}[ht]%
    \centering
    \includegraphics[width=\textwidth]{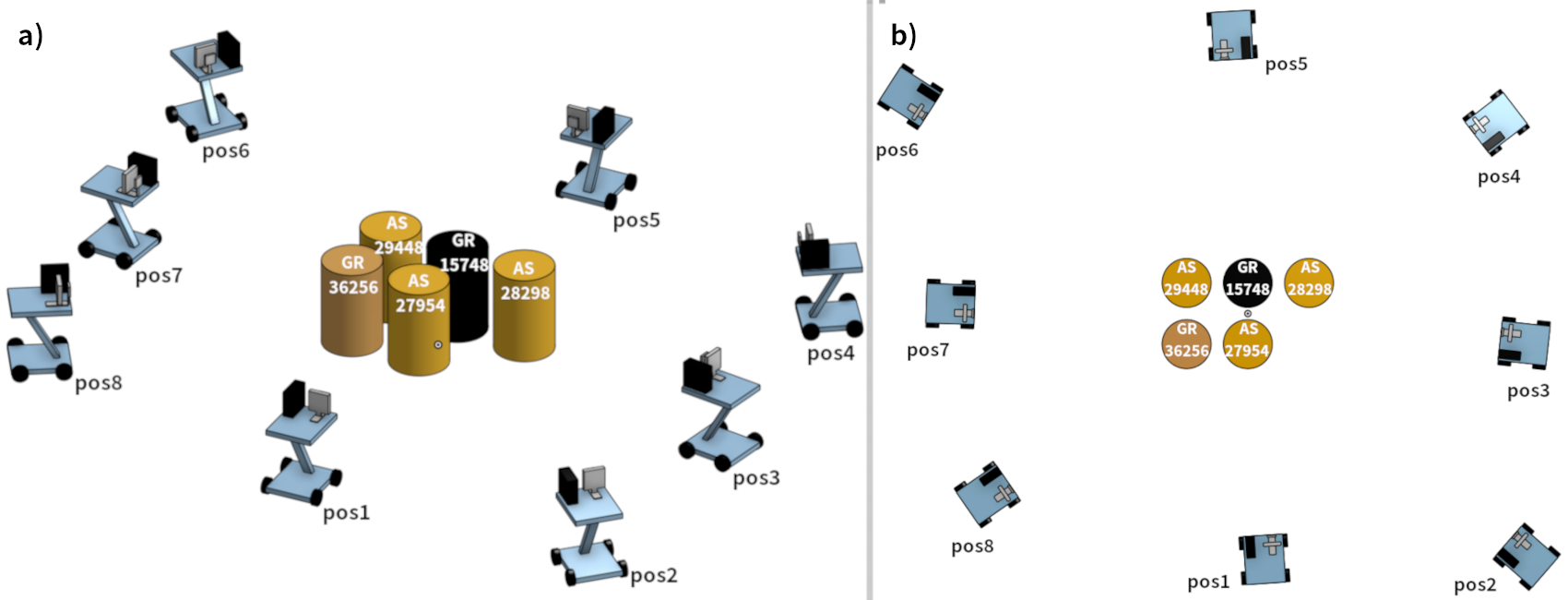}
    \caption{Schematic (CAD) view of the barrel configuration B and the measured poses. a) Isometric- b) top-view.} 
    \label{fig:cadArrangementB}
\end{figure}

\myfigref{fig:cadArrangementA} and \myfigref{fig:cadArrangementB} show the schematic views of the two measured configurations along with the corresponding poses. Both barrel ensembles consist of the same five barrels labeled as AS28298, AS29448, GR36256, AS27954 and GR15748. The only distinction between them lies in the exchange of positions of the AS28298 and AS29448 drums ($i$), as well as the movement of the GR15748 barrel to another position ($ii$).

\begin{figure}[ht]%
    \centering
    \includegraphics[width=\textwidth]{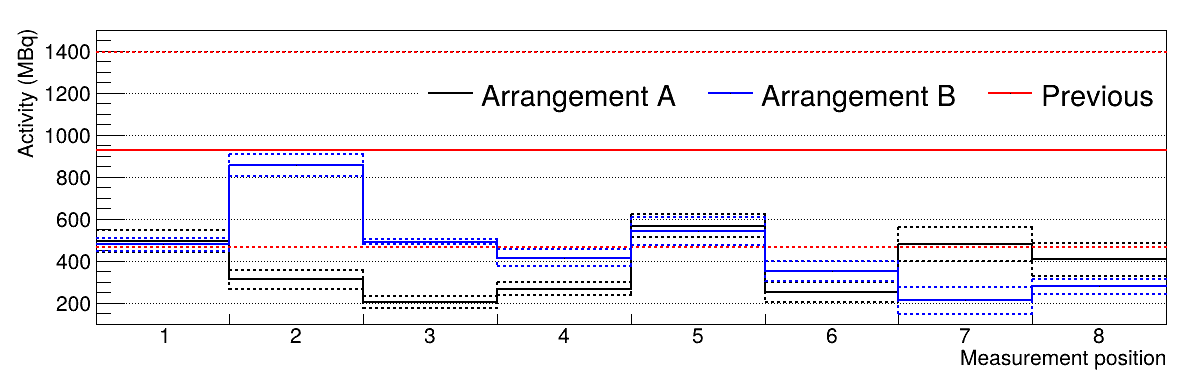}
    \caption{Average activity estimated for the two barrel configurations as a function of the imager pose. The value displayed at each pose corresponds to the average activity value obtained from all five PSDs in the imager.}\label{fig:activityComparisonAB}
\end{figure}

These two configurations were intended to explore and demonstrate the sensitivity of the hybrid imaging system for detecting and monitoring variations in the nuclear inspection of a complex scenario made of 5 different drums. To this aim the imaging system needs to have a relatively large field of view and a sufficiently large image resolving power. However, before applying CPU-demanding image reconstruction algorithms, the first quick real-time test that can be conducted during the nuclear inspection concerns the apparent surface activity measured for the two configurations, in a similar fashion as discussed before for the single barrel in~\mysecref{subsec:singlebarrel}. Here one is taking advantage of the large efficiency for single $\gamma$-ray detection with the five large uncollimated LaCl$_3$(Ce) crystals embedded in the Compton camera (see Sec.\ref{sec:setup}). ~\myfigref{fig:activityComparisonAB} shows the overall apparent activity determined for each pose. The very first remarkable aspect are the strong differences appreciated for some specific poses, despite the fact that both configurations comprise the same barrel units. Therefore, activity measurements can quickly and very sensitively uncover variations in the configuration of a certain barrel ensemble. On the other hand, it turns out that a proper interpretation solely based on the measured activities is not straight forward. As discussed below, these variations are mainly due to the different self-shielding effects ascribed to each configuration. 

A more detailed insight in \myfigref{fig:activityComparisonAB} shows the presence of two different trends in the measured data. On the one hand, for poses 1 and 5, both A and B configurations yield a very similar result in terms of apparent surface activity. On the other hand, differences between a factor of 2 (pose 7) and 3 (pose 2) are found for the remaining positions.
At this stage, for the interpretation of these results one needs to consider two aspects. Firstly, the rather different radiation absorption coefficient for GR barrels in comparison to AS, which is a factor 2 higher for the former. Secondly, the comparable surface activity levels of all barrels (Table~\ref{tab:activityBarrels}), which varies within a factor 2-3 with respect to their average value (126 MBq). It is worth noting that such a factor of 2-3 is comparable to the uncertainty expected for the implemented activity assessment methodology~\cite{Ridikas2005}. Thus, the results can be mainly explained in terms of shielding or radiation-absorption effects in each geometry or pose. Indeed, for each pose GR barrels are expected to induce a large shielding effect for the radiation coming from the barrel(s) behind them, and a corresponding decrease in the measured surface activity. On the contrary, AS-type barrels allow a much higher portion of high-energy $\gamma$-rays to pass through them. 
Therefore, the large difference in~\myfigref{fig:activityComparisonAB} for poses 2 and 3 is due to the fact that in configuraton A (see \myfigref{fig:cadArrangementA}) the barrel GR15748 is effectively shielding the imaging system from radiation arising from AS28298, whereas in configuration B for both poses 2 and 3 the imaging system is directly exposed to the radiation from AS28298 (see~\myfigref{fig:cadArrangementB}). 

The second largerst difference that can be observed in~\myfigref{fig:activityComparisonAB} between configurations A and B corresponds to pose 7. Again, as it can be inferred from~\myfigref{fig:cadArrangementA} and~\myfigref{fig:cadArrangementB}, the drum AS28298 in pose 7-A is directly exposed to the imager, whereas in pose 7-B it is significantly shielded by both GR-drums in the setup.

In summary, the aforediscussed apparent overall activity results reflect two important aspects. Firstly, significant differences in apparent surface activity can be observed for modifications of the barrel configuration. Secondly, due to activity self-shielding effects, the overall measured activity level and its interpretation is ambiguous without a detailed knowledge of the true activity in each barrel, its distribution inside the barrel and the filling material.

\begin{figure}[ht]%
    \centering
    \includegraphics[width=\textwidth]{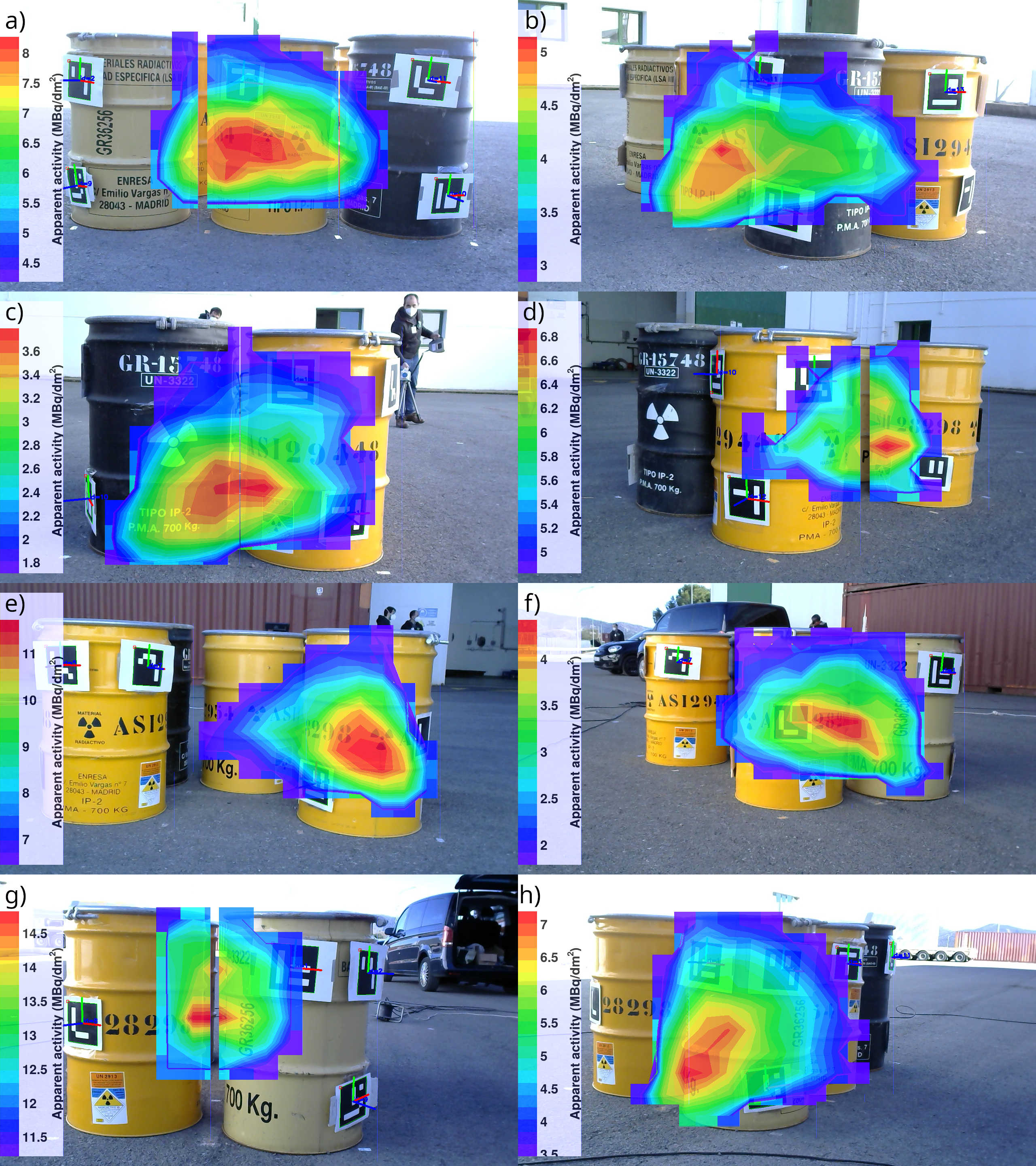}
    \caption{From a) to h), hybrid \gr and visible 2D images reconstructed for the eight poses in configuration A.}\label{fig:barrelArrangement02}
\end{figure}

\begin{figure}[ht]%
    \centering
    \includegraphics[width=\textwidth]{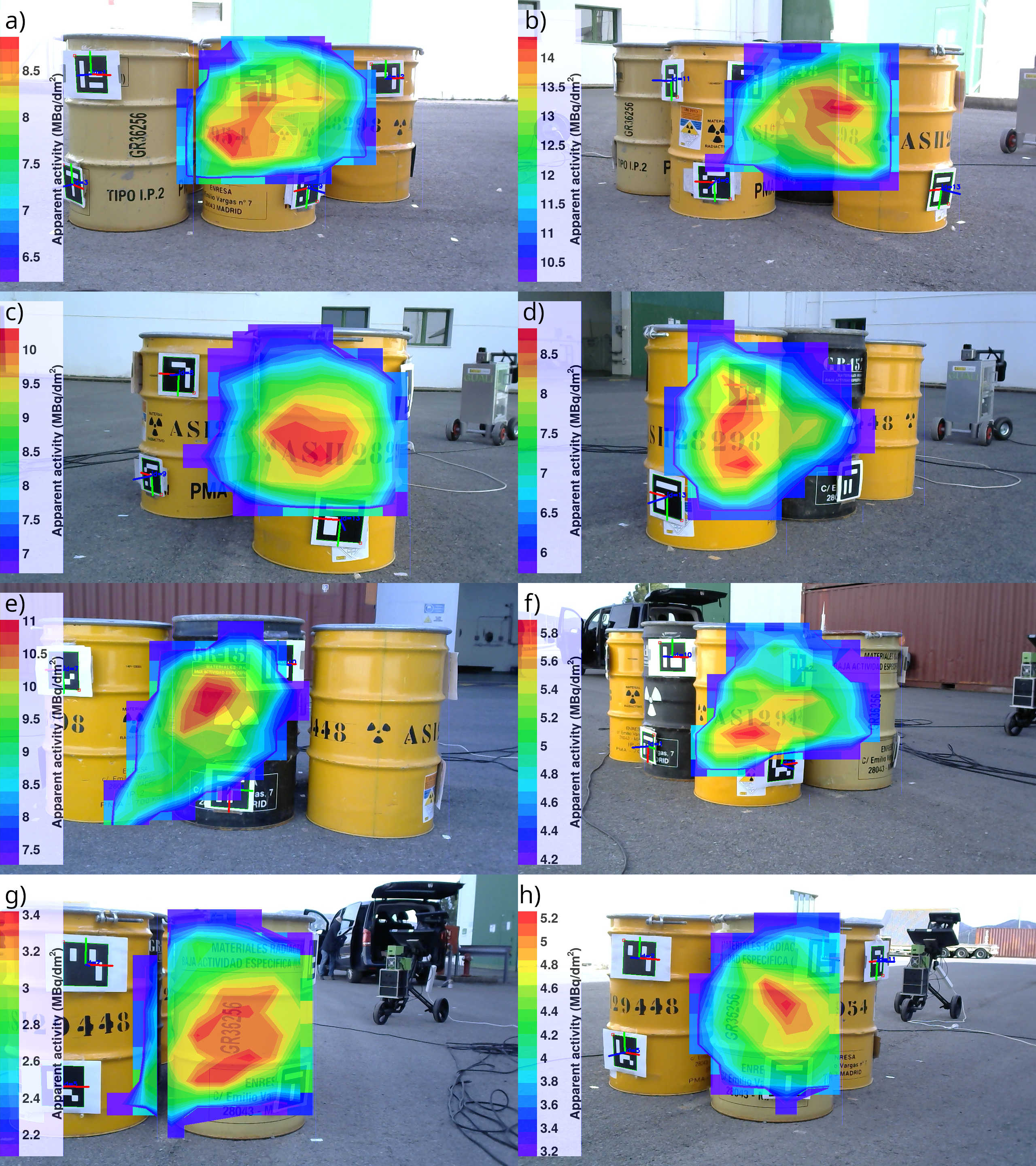}
    \caption{From a) to h), Hybrid \gr and visible 2D images reconstructed for the eight poses of configuration B.}\label{fig:barrelArrangement03a}
\end{figure}

Next, a brief discussion based on the 2D and 3D reconstructed images will be made for configurations A and B. \myfigref{fig:barrelArrangement02} and~\myfigref{fig:barrelArrangement03a} display the 2D reconstructed \gr and visible images for the drum ensembles corresponding to these two configurations. Despite the changes $i$ and $ii$, both sets of images indicate that the most active drum is AS28298. This statement becomes clear after inspecting the colour scale and numerical values depicted in poses 4 to 7 (\myfigref{fig:barrelArrangement02}-d to -g) and 2 to 4 (\myfigref{fig:barrelArrangement03a}-b to -d), for configurations A and B, respectively. However, the conventional CDR methodology (see~\mytabref{tab:activityBarrels}) yields an activity estimate, which is a factor of three larger for  AS27954 with respect to the one quoted for AS28298. This apparent contradiction may be ascribed, on one hand, to the rather conservative values that are commonly estimated with the conventional approach, where after several dose-rate measurements those with maximum values prevail in the averaged estimated quantity. This effect may be amplified by the fact that the activity distribution in AS28298 seems to be rather concentrated in a single central region of the barrel. The latter aspect can be better disentangled by comparing, for example, the relatively broad spatial distribution of low (up to 8 MBq) activity for AS27954 in pose 1 (\myfigref{fig:barrelArrangement02}-a), with respect to the rather high maximum (14.5 MBq) and narrow spot measured for AS28298 in pose 7 (\myfigref{fig:barrelArrangement02}-g).

\begin{figure}[ht]%
    \centering
    \includegraphics[width=\textwidth]{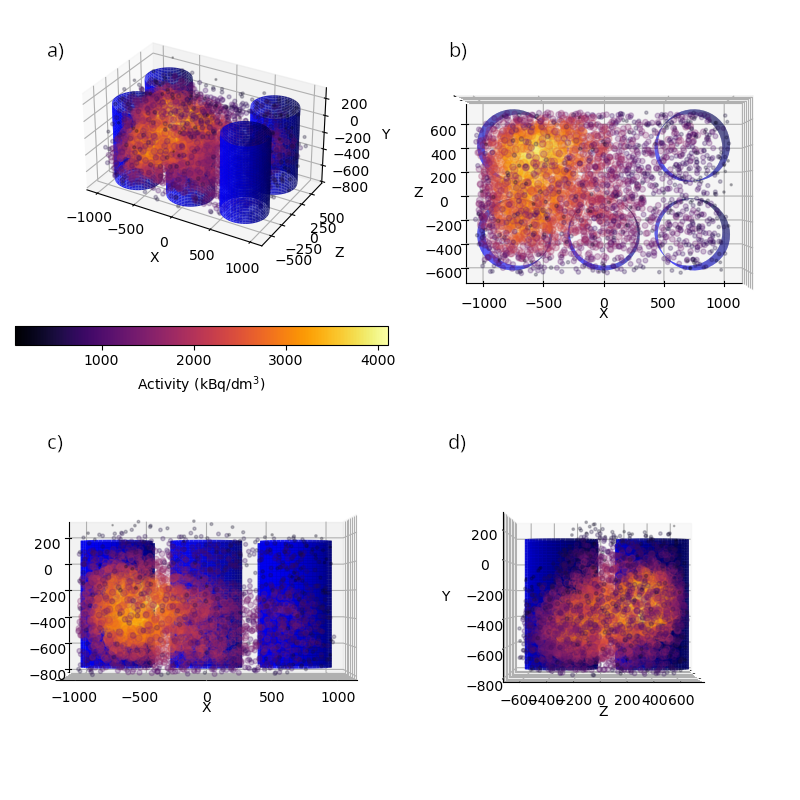}
    \caption{3D distribution of \gr radiation of the barrel arrangement A.}\label{fig:3Darrangement02}
\end{figure}

\begin{figure}[ht]%
    \centering
    \includegraphics[width=\textwidth]{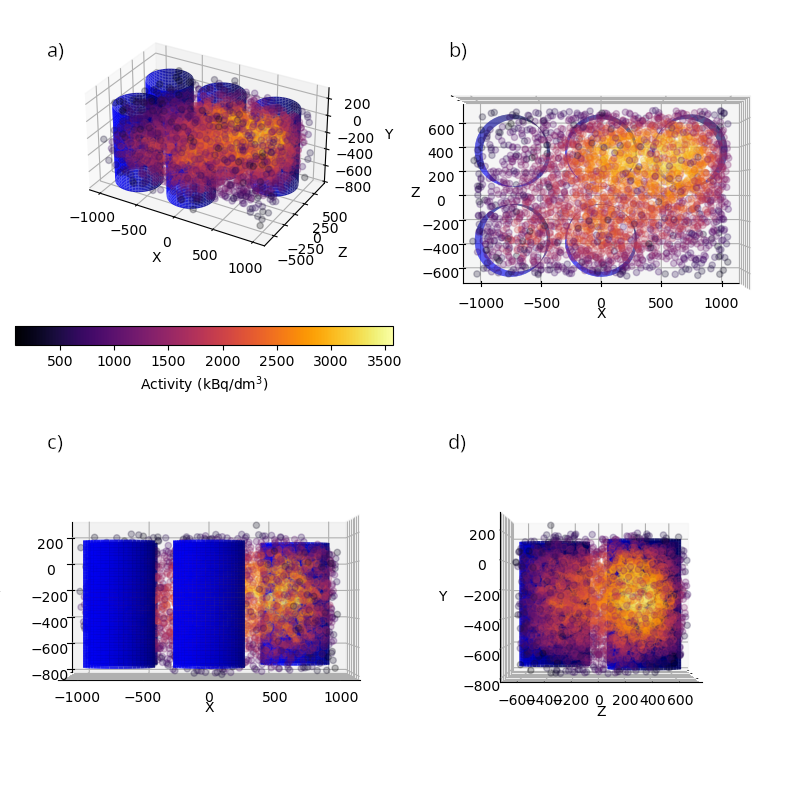}
    \caption{3D distribution of \gr radiation of the barrel arrangement C.}\label{fig:3Darrangement03a}
\end{figure}

Finally, the 3D reconstructed distributions are shown in  \myfigref{fig:3Darrangement02} and \myfigref{fig:3Darrangement03a} for barrel configurations A and B, respectively. An effective attenuation coefficient of 1 m$^{-1}$ was used for the image reconstruction. As explained in~\mysecref{subsubsec:3D}, this factor takes into account not only the attenuation of \grs in the drums but also a correction for the overlapping effect induced by the Compton arcs in the backprojection method. The 3D-reconstructed results obtained from these two independent set of measurements for each configuration are consistent with the previous interpretation that barrel AS28298 is the one showing largest overall surface activity.

\section{Conclusions and outlook}\label{sec:conclusions}
In this work we have shown how a portable high-efficiency $\gamma$-ray imager in conjunction with image-segmentation and computer vision techniques, allows for an efficient and advanced analysis of radioactive waste.
Demonstration measurements have been conducted at the disposal plant of El Cabril, which is managed by the national company ENRESA.
Once the system has been characterized and calibrated, the implemented methodology allows one to assess first, and in real time, the overall apparent activity level of the corresponding drum or object of interest, which is one of the first aspects of concern in any inspection of radioactive waste or materials. The overall apparent surface-activity obtained with the new system agrees, within a factor of $\sim$2, with previous estimates based on conventional dose-rate surface measurements.
An hybrid $\gamma$-ray and visual image can be produced also in real time, which allows one to reconstruct a 2D map of the apparent surface-activity distribution in the region of interest. Typical examination times for the examined activity levels and drum volumes are as short as 2 minutes for regular barrels with concrete or cellulose filling containing conventional LLW and ILW residues. Such a short measuring time, which is enabled by the high-efficiency design of the implemented Compton camera, is a significant advantage for this particular application because it may enable one to enhance the classification yield and reduce operator exposure.
Finally, we have also demonstrated how the combination of several measurements around a barrel, and around a barrel ensemble, allow one to make a quantitative and visual assessment of the different activity levels and distributions involved. For the latter aim both 2D-hybrid images and 3D images of the activity, have been reconstructed from a series of measurements carried out at the disposal plant of El Cabril, thereby utilizing two different scenarios with five different barrels.

A future improvement of the work presented here will involve a further optimisation towards real-time 2D and 3D image processing by leveraging GPU acceleration. This enhancement aims to provide operators with a comprehensive tool for rapidly assessing the activity level and homogeneity of radioactive materials, facilitating their effective management. 

We have also noticed that relatively complex barrel ensembles are subject to pronounced self-shielding and activity summation effects, which may challenge accuracy and resolution with state of the art \gr image and activity reconstruction methodologies. For this reason, in the near future we plan to perform dedicated measurements in order to disentangle the interplay between these effects, and to evaluate in a more detailed fashion the inherent systematic uncertainties, especially for complex classification environments.
Finally, we also envisage the future implementation of the recently developed gamma-neutron vision system, GNVision\cite{Lerendegui2024b,Lerendegui24, Patent}, as a potential enhancement towards a more complete and reliable inspection of radioactive waste.

\section*{Acknowledgments}
The corresponding author (VB) is a beneficiary of the Margarita Salas grant (MS21-178) for the requalification of the Spanish university system from the Ministry of Universities of the Government of Spain, financed by the European Union, NextGenerationUE. CDP and JLL acknowledge funding from Convenio CSIC-ENRESA "Desarrollo de un dispositivo para la identificación, cuantificación y distribución espacial de isótopos emisores gamma", 2014-2018. This work was also supported by the Spanish Ministerio de Ciencia e Innovaci\'on under grants PID2022-138297NB-C21, PID2019-104714GB-C21, FPA2017-83946-C2-1-P and CSIC for funding PIE-201750I26. Part of the hardware and software utilized in this work has been developed in the framework of the preceding HYMNS ERC-CoG Grant (Agreement Nr.681740) led by CDP. The authors acknowledges the support provided by postdoctoral grants FJC2020-044688-I and ICJ220-045122-I, funded by MCIN/AEI/10.13039/501100011033. The authors acknowledge discussions with S. Tortajada, F. Albiol and L. Caballero during the field-measurements at the disposal-plant of El Cabril.

\bibliographystyle{plain}
\bibliography{main}

\end{document}